\documentclass[multphys,vecphys]{svmult}

\usepackage{makeidx}         
\usepackage{graphicx}        
\usepackage{multicol}        

\makeindex             

\newcommand{\be}{\begin{equation}}
\newcommand{\ee}{\end{equation}}
\newcommand{\ba}{\begin{eqnarray}}
\newcommand{\ea}{\end{eqnarray}}

\newcommand{\nn}{\nonumber \\}
\newcommand{\nnb}{\begin{displaymath}}
\newcommand{\nne}{\end{displaymath}}

\newcommand{\x}{{\bf x}}
\renewcommand{\r}{\mbox{\boldmath $r$}}

\newcommand{\s}{{\bf s}}

\renewcommand{\E}{\mbox{\boldmath $E$}}

\newcommand{\ie}{{\em i.e.}}

\newcommand{\amp}{\vert \delta_{\bf k} \vert}
\newcommand{\ampp}{\vert \delta_{\bf k'} \vert}
\newcommand{\amppp}{\vert \delta_{\bf k''} \vert}
\newcommand{\ampppp}{\vert \delta_{\bf k'''} \vert}
\newcommand{\phik}{\phi_{\bf k}}
\newcommand{\phikp}{\phi_{\bf k'}}
\newcommand{\phikpp}{\phi_{\bf k''}}
\newcommand{\phikppp}{\phi_{\bf k'''}}

\def\simless{\mathbin{\lower 3pt\hbox
   {$\rlap{\raise 5pt\hbox{$\char'074$}}\mathchar"7218$}}}

\begin{document}

\title*{Phase Correlations and Topological Measures of Large-scale Structure}
\author{Peter Coles}
\institute{School of Physics \& Astronomy, \\ University of
Nottingham, \\ University Park, \\ Nottingham NG7 2RD, \\
 United
Kingdom \\
\texttt{Peter.Coles@Nottingham.ac.uk} }
%
%
\maketitle

\section{Introduction}

The process of gravitational instability initiated by small
primordial density perturbations is a vital ingredient of
cosmological models that attempt to explain how galaxies and
large-scale structure formed in the Universe. In the standard
picture (the ``concordance'' model), a period of accelerated
expansion (``inflation'') generated density fluctuations with
simple statistical properties through quantum processes
(Starobinsky 1979, 1980, 1982; Guth 1981; Guth \& Pi 1982;
Albrecht \& Steinhardt 1982; Linde 1982). In this scenario the
primordial density field is assumed to form a statistically
homogenous and isotropic Gaussian Random Field (GRF). Over years
of observational scrutiny this paradigm has strengthened its hold
in the minds of cosmologists and has survived many tests,
culminating in those furnished by the Wilkinson Microwave
Anisotropy Probe (WMAP; Bennett et al. 2003; Hinshaw et al 2003).

Gaussian random fields are the simplest fully-defined stochastic
processes (Adler 1981; Bardeen et al. 1986), which makes their
analysis relatively straightforward. Robust and powerful
statistical descriptors can be constructed that have a firm
mathematical underpinning and are relatively simple to implement.
Second-order statistics such as the ubiquitous power-spectrum
(e.g. Peacock \& Dodds 1996) furnish a complete description of
Gaussian fields. They have consequently yielded invaluable
insights into the behaviour of large-scale structure in the latest
generation of redshift surveys, such as the 2dFGRS (Percival et
al. 2001). Important though these methods undoubtedly are, the era
of precision cosmology we are now entering requires more thought
to be given to methods for both detecting and exploiting
departures from Gaussian behaviour.

Even if the primordial density fluctuations were indeed Gaussian,
the later stages of gravitational clustering must induce some form
of non-linearity. One particular way of looking at this issue is
to study the behaviour of Fourier modes of the cosmological
density field. If the hypothesis of primordial Gaussianity is
correct then these modes began with random spatial phases. In the
early stages of evolution, the plane-wave components of the
density evolve independently like linear waves on the surface of
deep water. As the structures grow in mass, they interact with
other in non-linear ways, more like waves breaking in shallow
water. These mode-mode interactions lead to the generation of
coupled phases. While the Fourier phases of a Gaussian field
contain no information (they are random), non-linearity generates
non-random phases that contain much information about the spatial
pattern of the fluctuations. Although the  significance of phase
information in cosmology is still not fully understood, there have
been a number of attempts to gain quantitative insight into the
behaviour of phases in gravitational systems. Ryden \& Gramann
(1991), Soda \& Suto (1992) and Jain \& Bertschinger (1998)
concentrated on the evolution of phase shifts for individual modes
using perturbation theory and numerical simulations. An
alternative approach was adopted by Scherrer, Melott \& Shandarin
(1991), who developed a practical method for measuring the phase
coupling in random fields that could be applied to real data. Most
recently Chiang \& Coles (2000), Coles \& Chiang (2000), Chiang
(2001) and Chiang, Naselsky \& Coles (2002) have explored the
evolution of phase information in some detail.

Despite this recent progress, there is still no clear
understanding of how the behaviour of the Fourier phases manifests
itself in more orthodox statistical descriptors.  In particular
there is much interest in the usefulness of the simplest possible
generalisation of the (second-order) power-spectrum, i.e. the
(third-order) bispectrum (Peebles 1980; Scoccimarro et al. 1998;
Scoccimarro, Couchman \& Frieman 1999; Verde et al. 2000; Verde et
al. 2001; Verde et al. 2002). Since the bispectrum is identically
zero for a Gaussian random field, it is generally accepted that
the bispectrum encodes some form of phase information but it has
never been elucidated exactly what form of correlation it
measures. Further possible generalisations of the bispectrum are
usually called polyspectra; they include the (fourth-order)
trispectrum (Verde \& Heavens 2001) or a related but simpler
statistic called the second-spectrum (Stirling \& Peacock 1996).
Exploring the connection between polyspectra and non-linearly
induced phase association is one of the  aims of this paper.

Gravitational instability is expected to generate phase
correlations (and non--Gaussianity) even if the primordial
fluctuations were Gaussian. The Cosmic Microwave Background (CMB)
allows us to probe the fluctuations while they are still in the
linear regime and thus test the level of primordial
non-Gaussianity without having to worry about non-linear effects.
A second aim of this paper is to explain how one can use phase
correlations in spherical harmonic expansions of temperature
fluctuations in order to detect departures from standard
fluctuation statistics.

Finally I discuss the use of topological invariants such as the
Euler--Poincar\'{e} characteristic of isodensity contours to
assess the level of non--Gaussianity in large-scale structure.

\section{Basic Statistical Tools}

\label{background}

I start by giving some general definitions of concepts which I
will later use in relation to the particular case of cosmological
density fields. In order to put our results in a clear context, I
develop the basic statistical description of cosmological density
fields; see also, e.g., Peebles (1980) and Coles \& Lucchin
(2002).

\subsection{Fourier Description}

I follow standard practice and consider a region of the Universe
having volume $V_u$, for convenience assumed to be a cube of side
$L\gg l_s$, where $l_s$ is the maximum scale at which there is
significant structure due to the perturbations. The region $V_u$
can be thought of as a ``fair sample'' of the Universe if this is
the case. It is possible to construct, formally, a ``realisation''
of the Universe by dividing it into cells of volume $V_u$ with
periodic boundary conditions at the faces of each cube. This
device is often convenient, but in any case one often takes the
limit $V_u \rightarrow\infty$. Let us denote by $\bar{\rho}$ the
mean density in a volume $V_u$ and take $\rho({\bf x})$ to be the
density at a point in this region specified by the position vector
${\bf x}$ with respect to some arbitrary origin. As usual, the
fluctuation is defined to be
\be
\delta(\x)=[\rho (\x)-\bar{\rho}]/\bar{\rho}. \ee We assume this
to be expressible as a Fourier series:
\be
\delta(\x)=\sum_{{\bf k}}\ \delta_{\bf k}\ \exp (i{\bf  k} \cdot
{\bf x})= \sum_{\bf k}\ \delta^*_{\bf k}\ \exp(-i{\bf k}\cdot {\bf
x}); \label{fourierseries} \ee the appropriate inverse
relationship is of the form \be  \delta_{\bf k} = {1 \over
V_u}\int_{V_u}\ \delta({\bf x}) \exp(-i{\bf k}\cdot {\bf x})d{\bf
x}.
 \ee The Fourier coefficients $\delta_{\bf
k}$ are complex quantities, \be \delta_{\bf k} = \amp
\exp{(i\phi_{\bf k})},\ee
 with amplitude $\amp$ and phase
$\phi_{\bf k}$. The assumption of periodic boundaries results in a
discrete ${\bf k}$-space representation; the sum is taken from the
Nyquist frequency $k_{\rm Ny}=2\pi/L$, where $V_u=L^3$, to
infinity. Note that as $L\rightarrow \infty$, $k_{\rm
Ny}\rightarrow 0$. Conservation of mass in $V_u$ implies
$\delta_{{\bf k}=0}=0$ and the reality of $\delta({\bf x})$
requires $\delta_{\bf k}^* = \delta_{-{\bf k}}$.

If, instead of the volume $V_u$, we had chosen a different volume
$V'_u$ the perturbation within the new volume would again be
represented by a series of the form (\ref{fourierseries}), but
with different coefficients $\delta_{\bf k}$.  Now consider a
(large) number $N$ of realisations of our periodic volume and
label these realisations by $V_{u1}$, $V_{u2}$, $V_{u3}$, ...,
$V_{uN}$. It is meaningful to consider the probability
distribution ${\cal P} (\delta_{\bf k})$ of the relevant
coefficients $\delta_{\bf k}$ from realisation to realisation
across this ensemble. One typically assumes that the distribution
is statistically homogeneous and isotropic, in order to satisfy
the Cosmological Principle, and that the real and imaginary parts
of $\delta_{\bf k}$ have a Gaussian distribution and are mutually
independent, so that
\be
{\cal P }(w) = {V_u^{1/2} \over (2\pi \alpha^2_k)^{1/2}} \exp
\Bigl (-{w^2 V_u \over 2 \alpha^2_k }\Bigr), \ee where $w$ stands
for either ${\rm Re}~[\delta_{\bf k}]$ or ${\rm Im}~[\delta_{\bf
k}]$ and $\alpha_k^2 = \sigma^2_k/2$; $\sigma_k^{2}$ is the
spectrum. This is the same as the assumption that the phases
$\phi_{\bf k}$ in equation (5) are mutually independent and
randomly distributed over the interval between $\phi=0$ and
$\phi=2\pi$. In this case the moduli of the Fourier amplitudes
have a Rayleigh distribution:
\be
{\cal P} (\vert \delta_{\bf k} \vert, \phik) d\vert \delta_{\bf
k}\vert d \phik = { \vert \delta_{\bf k}\vert V_u \over 2\pi
\sigma^2_k }\exp \Bigl(-{\vert \delta_{\bf k}\vert ^2 V_u \over
2\sigma_k^2 } \Bigr) d\vert \delta_{\bf k}\vert d \phik . \ee
Because of the assumption of statistical homogeneity and isotropy,
the quantity ${\cal P}(\delta_{\bf k})$ depends only on the
modulus of the wavevector ${\bf k}$ and not on its direction. It
is fairly simple to show that, if the Fourier quantities $\vert
\delta_{\bf k} \vert $ have the Rayleigh distribution, then the
probability distribution ${\cal P}(\delta)$ of $\delta=\delta({\bf
x})$ in real space is Gaussian, so that: \be {\cal P}(\delta)
d\delta = {1 \over (2\pi\sigma^2)^{1/2}} \exp
\Bigl(-{\delta^2\over2\sigma^2}\Bigr)d\delta,
\label{1_pt_gaussian} \ee where $\sigma^2$ is the variance of the
density field $\delta({\bf x})$. This is a strict definition of
Gaussianity. However, Gaussian statistics do not always require
the distribution (7) for the Fourier component amplitudes.
According to its Fourier expansion, $\delta({\bf x})$ is simply a
sum over a large number of Fourier modes whose amplitudes are
drawn from some distribution. If the phases of each of these modes
are random, then the Central Limit Theorem will guarantee that the
resulting superposition will be close to a Gaussian if the number
of modes is large and the distribution of amplitudes has finite
variance. Such fields are called weakly Gaussian.

\subsection{Covariance Functions \& Probability Densities}

I now discuss the real-space statistical properties of spatial
perturbations in $\rho$. The \emph{covariance function} is defined
in terms of the density fluctuation by
\be
\xi({\bf r})={\langle[\rho({\bf x})-\bar{\rho}] [\rho({\bf x}+{\bf
r})-\bar{\rho}]\rangle\over \bar{\rho}^2}=\langle\delta({\bf
x})\delta ({\bf x}+{\bf r})\rangle. \label{2_pt_def} \ee The angle
brackets in this expression indicate two levels of averaging:
first a volume average over a representative patch of the universe
and second an average over different patches within the ensemble,
in the manner of \S 2.1.
Applying the Fourier machinery to equation (\ref{2_pt_def}) one
arrives at the {\it Wiener--Khintchin theorem}, relating the
covariance to the spectral density function or power spectrum,
$P(k)$:
\be
\xi({\bf r})= \sum_{\bf k} \langle\vert \delta_{\bf k}\vert
^2\rangle \exp(-i{\bf k}\cdot {\bf r}), \label{weiner_finite} \ee
which, in passing to the limit $V_u\rightarrow\infty$, becomes \be
\xi({\bf r})={1 \over (2\pi)^3}\int P(k) \exp(-i{\bf k}\cdot {\bf
r}) d{\bf k} . \label{weiner_cont} \ee Averaging equation
(\ref{weiner_finite}) over ${\bf r}$ gives
\be
\langle \xi({\bf r}) \rangle _{{\bf r}} = {1 \over V_u} \sum_{\bf
k} \langle\vert \delta_{\bf k}\vert ^2\rangle \int \exp(-i{\bf
k}\cdot {\bf r}) d {\bf r} = 0. \ee

The function $\xi({\bf r})$ is the \emph{two--point} covariance
function. In an analogous manner it is possible to define spatial
covariance functions for $N>2$ points. For example, the
three--point covariance function is \be \zeta(r,s)  = {\langle
[\rho({\bf x})- \bar{\rho}] [\rho({\bf x}+{\bf
r})-\bar{\rho}][\rho ({\bf x} + {\bf s})-\bar{\rho}]\rangle
\over\bar{\rho}^3} \label{3_pt_def} \ee which gives \be \zeta({\bf
r},{\bf s}) = \langle\delta({\bf x})\delta({\bf x}+ {\bf r})
\delta({\bf x} + {\bf s})\rangle, \ee where the spatial average is
taken over all the points ${\bf x}$ and over all directions of
${\bf r}$ and ${\bf s}$ such that $\vert {\bf r}-{\bf s}\vert =t$:
in other words, over all points defining a triangle with sides
$r$, $s$ and $t$. The generalisation of (\ref{3_pt_def}) to $N>3$
is obvious.

The covariance functions are related to the moments of the
probability distributions of $\delta({\bf x})$. If the
fluctuations form a Gaussian random field then the N-variate
distributions of the set $\delta_i \equiv \delta({\bf x}_i)$ are
just multivariate Gaussians of the form
\be
{\cal P}_N (\delta_1, ...,\delta_N) = {1 \over {(2 \pi)^{N/2}
({\rm det}~ C)^{1/2}}} \exp \Bigl( -{1\over 2} \sum_{i,j} \delta_i
~ C_{ij}^{-1} ~ \delta_j \Bigr). \ee The correlation matrix
$C_{ij}$ can be expressed in terms of the covariance function
\be
    C_{ij} = \langle \delta_i \delta_j \rangle = \xi({\bf r}_{ij}).
\label{corr_matrix} \ee It is convenient to go a stage further and
define the N-point {\it connected} covariance functions as the
part of the average $\langle \delta_i ... \delta_N \rangle$ that
is not expressible in terms of lower order functions e.g.
\be
    \langle \delta_1 \delta_2 \delta_3 \rangle
    = \langle\delta_1\rangle_c\langle\delta_2\delta_3\rangle_c
+\langle\delta_2\rangle_c\langle\delta_1\delta_3\rangle_c
+\langle\delta_3\rangle_c \langle\delta_1\delta_2\rangle _c+
\langle \delta_1 \rangle_c\langle \delta_2
\rangle_c\langle\delta_3 \rangle_c + \langle
\delta_1\delta_2\delta_3\rangle_c, \ee where the connected parts
are $\langle \delta_1\delta_2\delta_3 \rangle_c$, $\langle
\delta_1 \delta_2\rangle_c$, etc.  Since $\langle \delta
\rangle=0$ by construction, $\langle \delta_1 \rangle_c = \langle
\delta_1\rangle=0$. Moreover, $\langle \delta_1 \delta_2 \rangle_c
= \langle \delta_1\delta_2 \rangle$ and $\langle \delta_1 \delta_2
\delta_3 \rangle_c = \langle \delta_1 \delta_2 \delta_3 \rangle$.
The second and third order connected parts are simply the same as
the covariance functions. Fourth and higher order quantities are
different, however. The connected functions are just the
multivariate generalisation of the cumulants $\kappa_N$ (Kendall
\& Stewart 1977).  One of the most important properties of
Gaussian fields is that all of their N-point connected covariances
are zero beyond N=2, so that their statistical properties are
fixed once the set of two--point covariances (\ref{corr_matrix})
is determined. All large-scale statistical properties are
therefore determined by the asymptotic behaviour of $\xi(r)$ as
$r\rightarrow \infty$.

\section{Phase Coupling} \label{phase_coupling}

In \S \ref{background} we pointed out that a convenient definition
of a Gaussian field could be made in terms of its Fourier phases,
which should by independent and uniformly distributed on the
interval $[0,2\pi]$. A breakdown of these conditions, such as the
correlation of phases of different wavemodes, is a signature that
the field has become non-Gaussian. In terms of cosmic large-scale
structure formation, non-Gaussian evolution of the density field
is symptomatic of the onset of non-linearity in the gravitational
collapse process, suggesting that phase evolution and non-linear

evolution are closely linked. A relatively simple picture emerges
for models where the primordial density fluctuations are Gaussian
and the initial phase distribution is uniform. When perturbations
remain small evolution proceeds linearly, individual modes grow
independently and the original random phase distribution is
preserved. However, as perturbations grow large their evolution
becomes non-linear and Fourier modes of different wavenumber begin
to couple together. This gives rise to phase association and
consequently to non-Gaussianity. It is clear that phase
associations of this type should be related in some way to the
existence of the higher order connected covariance functions,
which are traditionally associated with non-linearity and are
non-zero only for non-Gaussian fields. In this sections such a
relationship is explored in detail using an analytical model for
the non-linearly evolving density fluctuation field. Phase
correlations of a particular form are identified and their
connection to the covariance functions is established.

A graphic demonstration of the importance of phases in patterns
generally is given in Figure 1.
\begin{figure}
\centering
\includegraphics[width=0.45\textwidth]{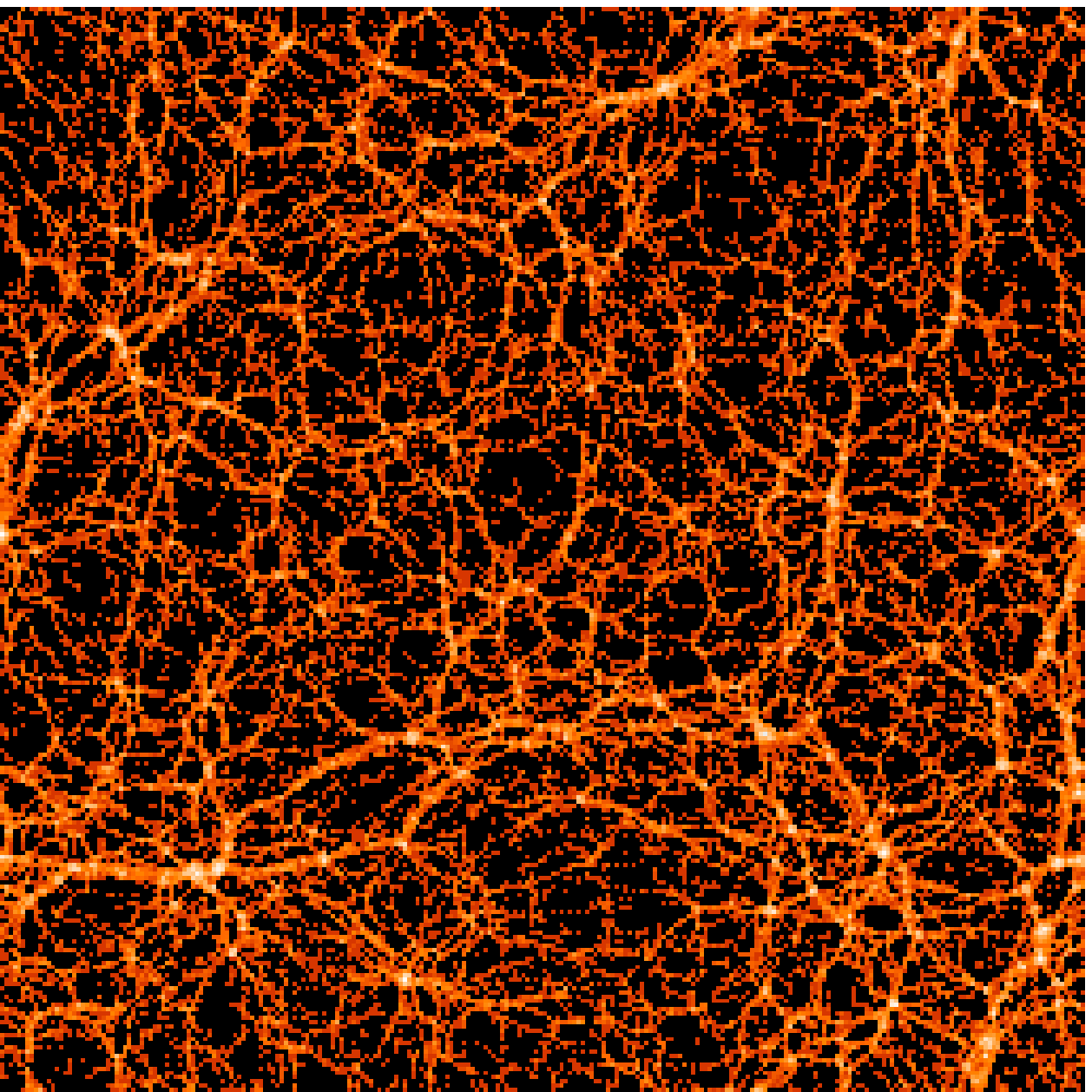}
\includegraphics[width=0.45\textwidth]{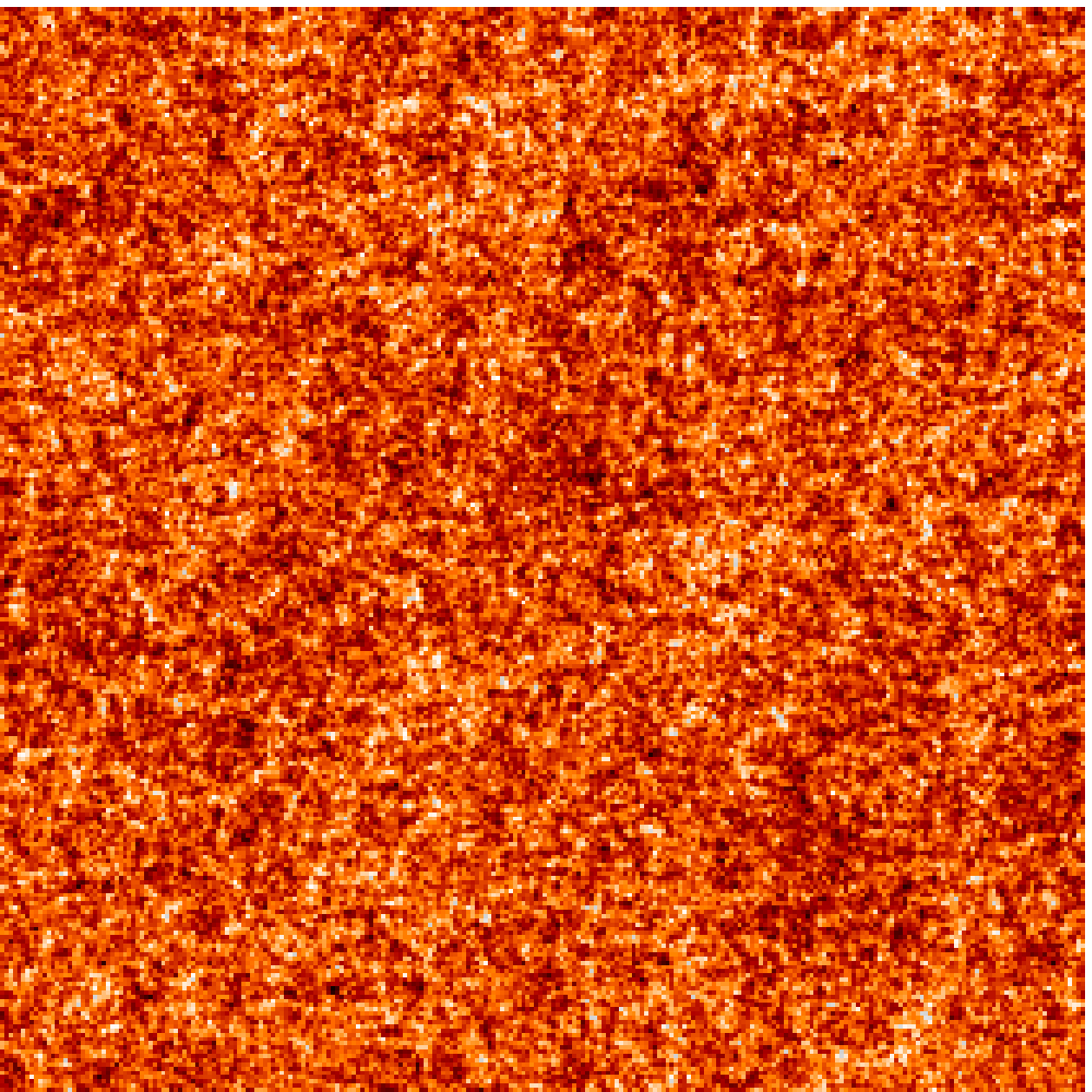}
\caption[]{Numerical simulation of galaxy clustering (left)
together with a version generated  randomly reshuffling the phases
between Fourier modes of the original picture (right).}
\label{eps1}
\end{figure}
Since the amplitude of each Fourier mode is unchanged in the phase
reshuffling operation, these two pictures have exactly the same
power-spectrum, $P(k)\propto|\tilde{\delta}({\bf k})|^2$. In fact,
they have more than that: they have exactly the same amplitudes
for all ${\bf k}$. They also have totally different morphology.
Further demonstrations of the importance of Fourier phases in
defining clustering morphology are given by Chiang (2001).

\subsection{Quadratic density fields}

It is useful at this stage to a particular form of non-Gaussian
field that serves both as a kind of phenomenological paradigm and
as a reasonably realistic model of non-linear evolution from
Gaussian initial conditions. The model involves a field which is
generated by a simple quadratic transformation of a Gaussian
distribution, hence the term {\em quadratic} non-linearity.
Quadratic fields have been discussed before from a number of
contexts (e.g. Coles \& Barrow 1987; Moscardini et al. 1991; Falk,
Rangarajan \& Srednicki 1993; Luo \& Schramm 1993; Luo 1994;
Gangui et al. 1994; Koyoma, Soda \& Taruya 1999; Peebles 1999a,b;
Matarrese, Verde \& Jimenez 2000; Verde et al. 2000; Verde et al.
2001; Komatsu \& Spergel 2001; Shandarin 2002; Bartolo, Matarrese
\& Riotto 2002); for further discussion see below. The motivation
is very similar to that of Coles \& Jones (1991), which introduced
the lognormal density field as an illustration of some of the
consequences of a more extreme form of non-linearity involving an
exponential transformation of the linear density field.

\subsection{A simple non-linear model}

We adopt a simple perturbative expansion of the form
\be
    \delta(\x) = \delta_1(\x) + \epsilon \delta_2(\x)
\label{nontransform} \ee  to mimic the non-linear evolution of the
density field. Although the equivalent transformation in formal
Eulerian perturbation theory is a good deal more complicated, the
kind of phase associations that we will deal with here are
precisely the same in either case. In terms of the Fourier modes,
in the continuum limit, we have for the first order Gaussian term
\be
    \delta_1(\x) = \int d^3 k \, \, \amp \exp{[i \phik]} \exp{[i{\bf k} \cdot\x]}
\ee and for the second-order perturbation
\be
    \delta_2(\x) = \left[\delta_1(\x)\right]^2  =
    \int d^3 k \, d^3 k' \, \, \amp \ampp
    \exp{[i(\phik + \phikp)]} \, \exp{[i({\bf k}+ {\bf k'})\cdot\r]}.
\label{quad_phase} \ee The quadratic field, $\delta_2$,
illustrates the idea of mode coupling associated with non-linear
evolution. The non-linear field depends on a specific harmonic
relationship between the wavenumber and phase of the modes at
${\bf k}$ and ${\bf k'}$. This relationship between the phases in
the non-linear field, i.e. \be \phi_{\bf k} + \phi_{{\bf k'}} =
\phi_{{\bf k}+ {\bf k'}}, \label{qpc} \ee where the RHS represents
the phase of the non-linear field, is termed {\em quadratic} phase
coupling.

\subsection{The two-point covariance function}

The two-point covariance function can be calculated using the
definitions of \S \ref{background}, namely
\be
    \xi(r) = \langle \delta(\x) \delta(\x+\r) \rangle.
\label{2point} \ee Substituting the non-linear transform for
$\delta(\x)$ (equation \ref{nontransform}) into this expression
gives four terms
\be
    \xi(r) = \langle \delta_1(\x) \delta_1(\x+\r) \rangle +
    \epsilon \langle \delta_1(\x) \delta_2(\x+\r) \rangle +
    \epsilon \langle \delta_2(\x) \delta_1(\x+\r) \rangle +
    \epsilon^2 \langle \delta_2(\x) \delta_2(\x+\r) \rangle.
\label{expand2pt} \ee The first of these terms is the linear
contribution to the covariance function whereas the remaining
three give the non-linear corrections. We shall focus on the
lowest order term for now.

As we outlined in Section 2, the angle brackets $\langle \rangle$
in these expressions are expectation values, formally denoting an
average over the probability distribution of $\delta(\x)$.
Under the fair sample hypothesis we replace the expectation values
in equation (\ref{2point}) with averages over a selection of
independent volumes so that $ \langle \rangle \rightarrow \langle
\rangle_{\mbox{\small{vol, real}}}$. The first average is simply a
volume integral over a sufficiently large patch of the universe.
The second average is over various realisations of the $\delta_k$
and $\phi_k$ in the different patches. Applying these rules to the
first term of equation (\ref{expand2pt}) and performing the volume
integration gives
\be
    \xi_{11}(r) =  \int d^3 k \, d^3 k' \, \, \langle \amp \ampp
    \exp{[i(\phik + \phikp)]} \rangle_{\mbox{\small{real}}}\,
    \delta_D({\bf k}+ {\bf k'}) \exp{[i{\bf k'}\cdot\s]},
\label{realav} \ee where $\delta_D$ is the Dirac delta function.
The above expression is simplified  given the reality condition
\be
    \delta_{\bf k} = \delta^*_{-\bf k},
\label{reality} \ee from which it is evident that the phases obey
\be
    \phi_{\bf k} + \phi_{-\bf k} = 0\, \, \, \mbox{mod}[2\pi].
\label{phase_property} \ee Integrating equation (\ref{realav}) one
therefore finds that
\be
    \xi_{11}(r) =  \int d^3 k \, \, \langle \amp^2 \rangle_{\mbox{\small{real}}}
    \exp{[- i {\bf k}\cdot\s]}.
\ee so that the final result is independent of the phases. Indeed
this is just the Fourier transform relation between the two-point
covariance function and the power spectrum we derived in \S 2.1.

\subsection{The three-point covariance function}

Using the same arguments outlined above it is possible to
calculate the 3-point connected covariance function, which is
defined as
\be
    \zeta(\r,\s) = \langle \delta(\x) \delta(\x+\r)
\delta(\x+\s) \rangle_c. \ee Making the non-linear transform of
equation (\ref{nontransform}) one finds the following
contributions \ba
    \zeta(\r,\s) & = & \langle \delta_1(\x) \delta_1(\x+\r) \delta_1(\x+\s)\rangle_c +
    \epsilon \langle \delta_1(\x) \delta_1(\x+\r)
\delta_2(\x+\s)\rangle_c  \nn & & + \mbox{perms}(121,211)
    + \epsilon^2 \langle \delta_1(\x) \delta_2(\x+\r)
\delta_2(\x+\s) \rangle_c \nn & & + \mbox{perms}(212,221)+
\epsilon^3 \langle \delta_2(\x) \delta_2(\x+\r)
\delta(\x+\s)\rangle_c. \label{expand3pt} \ea Again we consider
first the lowest order term. Expanding in terms of the Fourier
modes and once again replacing averages as prescribed by the fair
sample hypothesis gives \ba \zeta_{111}(\r,\s) & = & \int d^3 k \,
d^3 k' \, d^3 k''\, \, \, \langle \amp \ampp
        \amppp \exp{[i(\phik + \phikp + \phikpp)]}
    \rangle_{\mbox{\small{real}}}\nn
    & &    \times \delta_D({\bf k}+ {\bf k'} + {\bf k''}) \exp{[i{\bf k'}\cdot\r]} \exp{[i{\bf k''}\cdot\s]}.
\label{fourier3pt} \ea Recall that $\delta_1$ is a Gaussian field
so that $\phik$, $\phikp$ and $\phikpp$ are independent and
uniformly random on the interval $[0,2\pi]$. Upon integration over
one of the wavevectors the phase terms is modified so that its
argument contains the sum $(\phik + \phikp + \phi_{-{\bf k}-{\bf
k''}})$, or a permutation thereof. Whereas the reality condition
of equation (\ref{reality}) implies a relationship between phases
of anti-parallel wavevectors, no such conditions hold for modes
linked by the triangular constraint imposed by the Dirac delta
function. In other words, except for serendipity,
\be
    \phik + \phikp + \phi_{-{\bf k}-{\bf k''}} \neq 0.
\label{phase_constraint} \ee In fact due to the circularity of
phases, the resulting sum is still just uniformly random on the
interval $[0,2\pi]$ if the phases are random. Upon averaging over
sufficient realisations, the phase term will therefore cancel to
zero so that the lowest order contribution to the 3-point function
vanishes, \ie \ $\zeta_{111}(\r,\s) = 0$. This is not a new
result, but it does explicitly illustrate how the vanishing of the
three-point connected covariance function arises in terms of the
Fourier phases.

Next consider the first non-linear contribution to the 3-point
function given by
\be
    \zeta_{112}(\r,\s) = \epsilon \langle \delta_1(\x) \delta_1(\x+\r)
    \delta_2(\x+\s)\rangle,
\ee or one of its permutations. In this case one of the arguments
in the average is the field $\delta_2(\x)$, which exhibits
quadratic phase coupling of the form (\ref{qpc}). Expanding this
term to the point of equation (\ref{fourier3pt}) using the
definition (\ref{quad_phase}) one obtains \ba \zeta_{112}(\r,\s) &
= & \int d^3 k \, d^3 k' \, d^3 k'' \, d^3k''' \,
    \, \, \nn & & \langle \amp \ampp \amppp \ampppp
    \exp{[i(\phik + \phikp +
    \phikpp + \phikppp)]}  \rangle_{\mbox{\small{real}}} \nn
    & & \times \delta_D({\bf k}+ {\bf k'} + {\bf k''}+ {\bf k'''}) \nn & & \times \exp{[i{\bf k'}\cdot\r]}
    \exp{[i({\bf k''}+{\bf k'''})\cdot\s]}.
\label{fourier3ptnl} \ea Once again the Dirac delta function
imposes a general constraint upon the configuration of
wavevectors.  Integrating over one of the ${\bf k}$ gives ${\bf
k'''} = -{\bf k} -{\bf k'} -{\bf k''}$ for example, so that the
wavevectors must form a closed loop. This general constraint
however, does not specify a precise shape of loop, instead the
remaining integrals run over all of the different possibilities.
At this point we may constrain the problem more tightly by noting
that most combinations of the ${\bf k}$ will contribute zero to
$\zeta_{(112)}$. This is because of the circularity property of
the phases and equation (\ref{phase_constraint}). Indeed, the only
nonzero contributions arise where we are able to apply the phase
relation obtained from the reality constraint, equation
(\ref{phase_property}). In other words the properties of the
phases dictate that the wavevectors must align in anti-parallel
pairs: ${\bf k}= -{\bf k'}$, ${\bf k''} = -{\bf k'''}$ and so
forth.

There is a final constraint that must be imposed upon the ${\bf
k}$ if $\zeta$ is the {\em connected} $3$-point covariance
function. In a graph theoretic sense, the general (unconnected)
$N$-point function $ \langle \delta_{l_1}(\x_1) \delta_{l_2}(\x_2)
... \delta_{l_N}(\x_N)\rangle $ can be represented geometrically
by a sum of tree diagrams.  Each diagram consists of $N$ nodes of
order $l_i$, representing the $\delta_{l_i}(\x_i)$, and a number
of linking lines denoting their correlations; see Fry (1984) or
Bernardeau (1992) for more detailed accounts. Every node is made
up of $l_i$ internal points, which represent a factor $\delta_{\bf
k } = \amp \exp{(i\phik)}$ in the Fourier expansion. According to
the rules for constructing diagrams, linking lines may join one
internal point to a single other, either within the same node or
in an external node. The {\em connected} covariance functions are
represented specifically by the subset of diagrams for which every
node is linked to at least one other, leaving none completely
isolated. This constraint implies that certain pairings of
wavevectors do not contribute to the connected covariance
function. For more details, see Watts \& Coles (2002).

The above constraints may be inserted into equation
(\ref{fourier3ptnl}) by re-writing the Dirac delta function as a
product over Delta functions of two arguments, appropriately
normalised. There are only two allowed combinations of wavevectors
so we have
\be
    \delta_D({\bf k}+{\bf k'}+{\bf k''}+{\bf k'''}) \rightarrow
\frac{1}{2V_u}[\delta_D({\bf k}+{\bf k''})\delta_D({\bf k'''}+{\bf
k'''})+\delta_D({\bf k}+{\bf k'''})\delta_D({\bf k'}+{\bf k''})].
\ee Integrating over two of the ${\bf k}$ and using equation
(\ref{phase_property}) eliminates the phase terms and leaves the
final result
\be
\zeta_{112}(\r,\s) = \frac{1}{V_u}\int d^3 k \, d^3 k'
    \, \, \langle \amp^2 \ampp^2
    \rangle_{\mbox{\small{real}}}
    \exp{[i{\bf k'}\cdot\r]} \exp{[-i({\bf k}+{\bf k'})\cdot\s]}.
\label{3pt_final} \ee The existence of this quantity has therefore
been shown to depend on the quadratic phase coupling of Fourier
modes. The relationship between modes and the interpretation of
the tree diagrams is also dictated by the properties of the
phases.

One may apply the same rules to the higher order terms in equation
(\ref{expand3pt}). It is immediately clear that the $\zeta_{122}$
terms are zero because there is no way to eliminate the phase term
$\exp{[i(\phik + \phikp + \phikpp + \phikppp + \phi_{{\bf
k''''}})]}$, a consequence of the property equation
(\ref{phase_constraint}). Diagrammatically this corresponds to an
unpaired {\em internal} point within one of the nodes of the tree.
The final, highest order contribution to the 3-point function is
found to be \ba \zeta_{222}(\r,\s) & = & \frac{1}{V_u^2}\int d^3 k
\, d^3 k' \, d^3 k''
    \, \, \langle \amp^2 \ampp^2 \amppp^2
    \rangle_{\mbox{\small{real}}}\nn
    & & \times    \exp{[i({\bf k}-{\bf k'})\cdot\r]} \exp{[i({\bf k'}-{\bf k''})\cdot\s]},
\ea where the phase and geometric constraints allow 12 possible
combinations of wavevectors.

\subsection{Power-spectrum and Bispectrum}
The formal development of the relationship between covariance
functions and power-spectra developed above suggests the
usefulness of higher--order versions of $P(k)$. It is clear from
the above arguments  that a more convenient notation for the
power-spectrum than that introduced in \S 2.1 is
\be
\langle \delta_{{\bf k}} \delta_{{\bf k'}} \rangle = (2\pi)^3 P(k)
\delta_D({\bf k}+{\bf k'}). \ee The connection between phases and
higher-order covariance functions obtained above also suggests
defining higher-order polyspectra of the form
\be
\langle \delta_{{\bf k}} \delta_{{\bf k'}} \ldots \delta_{{\bf
k}^{(n)}} \rangle = (2\pi)^3 P_{n}({\bf k},{\bf k'},\ldots {\bf
k}^{(n)}) \delta_D({\bf k}+{\bf k'}+\ldots {\bf k}^{(n)})
\label{polysp} \ee where the occurrence of the delta-function in
this expression arises from a generalisation of the reality
constraint given in equation (\ref{phase_property}); see, e.g.,
Peebles (1980). Conventionally the version of this with $n=3$
produces the bispectrum, usually called $B({\bf k},{\bf k'},{\bf
k''})$ which has found much effective use in recent studies of
large-scale structure (Peebles 1980; Scoccimarro et al. 1998;
Scoccimarro, Couchman \& Frieman 1999; Verde et al. 2000; Verde et
al. 2001; Verde et al. 2002). It is straightforward to show that
the bispectrum is the Fourier-transform of the (reduced)
three--point covariance function by following similar arguments;
see, e.g., Peebles (1980).

Note that the delta-function constraint requires the bispectrum to
be zero except for $k$-vectors (${\bf k}$, ${\bf k'}$, ${\bf
k''}$) that form a triangle in $k$-space. It is clear that the
bispectrum can only be non-zero when there is a definite
relationship between the phases accompanying the modes whose
wave-vectors form a triangle. Moreover the pattern of phase
association necessary to produce a real and non-zero bispectrum is
precisely that which is generated by quadratic phase association.
This shows, in terms of phases, why it is that the leading order
contributions to the bispectrum emerge from second-order
fluctuations of a Gaussian random field. The bispectrum measures
quadratic phase coupling.

Three-point phase correlations have another interesting property.
While the bispectrum is usually taken to be an ensemble-averaged
quantity, as defined in equation (37), it is interesting to
consider products of terms $\delta_{{\bf k}} \delta_{{\bf k'}}
\delta_{{\bf k''}}$ obtained from an individual realisation.
According to the fair sample hypothesis discussed above we would
hope appropriate averages of such quantities would yield an
estimate of the bispectrum. Note that
\be
\delta_{{\bf k}} \delta_{{\bf k'}} \delta_{{\bf k''}} =
\delta_{{\bf k}} \delta_{{\bf k'}} \delta_{-{\bf k}- {\bf k'}} =
\delta_{{\bf k}} \delta_{{\bf k'}} \delta^*_{{\bf k}+ {\bf
k'}}\equiv \beta({\bf k}, {\bf k'}), \ee using the requirement
(\ref{phase_property}), together with the triangular constraint we
discussed above. Each $\beta({\bf k},{\bf k'})$ will carry its own
phase, say $\phi_{{\bf k},{\bf k'}}$, which obeys
\be
\phi_{{\bf k},{\bf k'}}=\phi_{{\bf k}}+\phi_{{\bf k'}}-\phi_{{\bf
k}+ {\bf k'}}. \label{phase_recon} \ee It is evident from this
that it is possible to recover the complete set of phases
$\phi_{{\bf k}}$ from the bispectral phases $\phi_{{\bf k},{\bf
k'}}$, up to a constant phase offset corresponding to a global
translation of the entire structure (Chiang \& Coles 2000). This
furnishes a conceptually simple method of recovering missing or
contaminated phase information in a consistent way, an idea which
has been exploited, for example, in speckle interferometry
(Lohmann, Weigelt \& Wirnitzer 1983). In the case of quadratic
phase coupling, described by equation (\ref{qpc}), the
left-hand-side of equation (\ref{phase_recon}) is identically zero
leading to a particularly simple approach to this problem.

\section{Phase Correlations in the CMB}

Since the release of the first (preliminary) WMAP data set it has
been subjected to a number of detailed independent analyses that
have revealed some surprising features. Eriksen et al. (2004) have
pointed out the existence of a North-South asymmetry suggesting
that the cosmic microwave background (CMB) revealed by the WMAP
data is not statistically homogeneous over the celestial sphere.
This is consistent with the results of Coles et al. (2004) who
found evidence for phase correlations in the WMAP data; see also
Hajian \& Souradeep (2003) and Hajian, Souradeep \& Cornish
(2004). The low--order multipoles of the CMB also display some
peculiarities (de Oliveira-Costa et al. 2004a; Efstathiou 2004).
Vielva et al. (2004) found significant non--Gaussian behaviour in
a wavelet analysis of the same data, as did Chiang et al. (2004),
Larson \& Wandelt (2004) and Park (2004). Other analyses of the
statistical properties of the WMAP have yielded results consistent
with the standard form of fluctuation statistics (Komatsu et al.
2003; Colley \& Gott 2003). These unusual properties may well be
generated by residual foreground contamination (Banday et al.
2003; Naselsky et al. 2003; de Oliveira-Costa et al. 2004; Dineen
\& Coles 2004) or other systematic effects, but may also provide
the first hints of physics beyond the standard cosmological model.

In order to tap the rich source of information provided by future
CMB maps it is important to devise as many independent statistical
methods as possible to detect, isolate and diagnose the various
possible causes of departures from standard statistics. One
particularly fruitful approach is to look at the behaviour of the
complex coefficients that arise in a spherical harmonic analysis
of CMB maps. Chiang et al. (2004), Chiang, Naselsky \& Coles
(2004), and Coles et al. (2004) have focussed on the phases of
these coefficients on the grounds that a property of a
statistically homogenous and isotropic GRF is that these phases
are random. Phases can also be use to test for the presence of
primordial magnetic fields (Chen et al. 2004; Naselsky et al.
2004) or evidence of non-trivial topology (Dineen, Rocha \& Coles
2004).

\subsection{Spherical Harmonics and Gaussian Fluctuations}

We can describe the distribution of fluctuations in the microwave
background over the celestial sphere using a sum over a set of
spherical harmonics:
\begin{equation}
\label{deltatovert} \Delta (\theta, \phi)= \frac{T(\theta ,\phi
)-\bar{T}}{\bar{T}}=\sum _{l=1}^{\infty }\sum _
{m=-l}^{m=+l}a_{l,m}Y_{lm}(\theta ,\phi ).
\end{equation}
Here $\Delta(\theta,\phi)$ is the departure of the temperature
from the average at angular position $(\theta ,\phi)$ on the
celestial sphere in some coordinate system, usually galactic. The
$Y_{lm}(\theta ,\phi)$ are spherical harmonic functions which we
define in terms of the Legendre polynomials $P_{lm}$ using
\begin{equation}
Y_{lm}(\theta ,\phi )= (-1)^m
\sqrt{\frac{(2l+1)(l-m)!}{4\pi(l+m)!}} P_{lm}(\cos \theta) {\rm e}
^{im\phi},
\end{equation}
i.e. we use the Condon-Shortley phase convention. In Equation (1),
the $a_{l,m}$ are complex coefficients which can be written
\begin{equation}
a_{l,m}=x_{l,m} + i y_{l,m} = |a_{l,m}|\exp[i\phi_{l,m}].
\end{equation}
Note that, since $\Delta$ is real, the definitions (40) \& (41)
requires the following relations between the real and imaginary
parts of the $a_{l,m}$: if $m$ is odd then
\begin{eqnarray}
x_{l,m}= \Re (a_{l,m}) &  = & - \Re(a_{l,-m})=-x_{l,-m},
\nonumber\\ y_{l,m}= \Im(a_{l,m}) &  = & \Im(a_{l,-m}) = y_{l,-m};
\end{eqnarray}
while if $m$ is even
\begin{eqnarray}
x_{l,m}=\Re (a_{l,m}) & = & \Re(a_{l,-m})=x_{l,-m}, \nonumber\\
y_{l,m}=\Im(a_{l,m}) & = & - \Im(a_{l,-m})=y_{l,-m};
\end{eqnarray}
and if $m$ is zero then
\begin{equation}
\Im(a_{l,m}) =y_{l,0} = 0.
\end{equation}
From this it is clear that the $m=0$ mode always has zero phase,
and there are consequently only $l$ independent phase angles
describing the harmonic modes at a given $l$. Without loss of
information we can therefore restrict our analysis to $m\geq 0$.

If the primordial density fluctuations form a Gaussian random
field in space the temperature variations induced across the sky
form a Gaussian random field over the celestial sphere. This means
that
\begin{equation}
\langle a_{l,m}a_{l',m'}^* \rangle = C_l \delta_{ll'}\delta_{mm'},
\end{equation}
where $C_l$ is the angular power spectrum, the subject of much
scrutiny in the context of the cosmic microwave background (e.g.
Hinshaw et al. 2003), and $\delta_{xx'}$ is the Kronecker delta
function. Since the phases are random, the stochastic properties
of a statistically homogeneous and isotropic Gaussian random field
are fully specified by the $C_l$, which determines the variance of
the real and imaginary parts of $a_{l,m}$ both of which are
Gaussian:
\begin{equation}
\sigma^{2} (x_{l,m}) = \sigma^{2}(y_{l,m})= \sigma_l^2=
\frac{1}{2} C_l.
\end{equation}

\subsection{Testing for Phase Correlations}

The approach we take  is to assume that we have available a set of
phases $\phi_{l,m}$ corresponding to a set of spherical harmonic
coefficients $a_{l,m}$ obtained from a data set, either real or
simulated. We can also form phase differences in according to
\begin{equation}
D_m(l)=\phi_{l,m+1}-\phi_{l,m}. \end{equation} If the orthodox
cosmological interpretation of temperature fluctuations is
correct, the phases of the $a_{l,m}$ should be random and so
should phase differences of the form $\phi_{l,m+1}-\phi_{l,m}$ and
$\phi_{l+1,m}-\phi_{l,m}$. Let us assume, therefore, that we have
$n$ generic angles, $\theta_1 \ldots \theta_n$. Under the standard
statistical assumption these should be random, apart from the
constraints described in the previous section. The first thing we
need is a way of testing whether a given set of phase angles is
consistent with being drawn from uniform distribution on the unit
circle. This is not quite as simple as it seems, particularly if
one does not want to assume any particular form for actual
distribution of angles, such as a bias in a particular direction;
see Fisher (1993). Fortunately, however, there is a fully
non--parametric method available, based on the theory of order
statistics, and known as as Kuiper's statistic (Kuiper 1960).

Kuiper's method revolves around the construction of a statistic,
$V$, obtained from the data via the following prescription. First
the angles are sorted into ascending order, to give the set
$\{\theta _{1},\ldots ,\theta _{n}\}$. It does not matter whether
the angles are defined to lie in $[0,2\pi]$, $[-\pi,+\pi]$ or
whatever. Each angle $\theta _{i}$ is divided by $2\pi$ to give a
set of variables $X_{i}$, where $i=1\ldots n$. From the set of
$X_i$ we derive two values $S^+_{n}$ and $S^-_{n}$ where
\begin{equation}
S^{+}_{n} = {\rm max}
\left\{\frac{1}{n}-X_{1},\frac{2}{n}-X_{2},\ldots ,1-X_{n}\right\}
\end{equation}
and   \begin{equation} S^{-}_{n} = {\rm max}
\left\{X_{1},X_{2}-\frac{1}{n},\ldots
\dot{,}X_{n}-\frac{n-1}{n}\right\}.
\end{equation}
Kuiper's statistic, $V$, is then defined as
\begin{equation}
\label{TestStatisticV} V=(S^{+}_{n}+S^{-}_{n})\cdot
\left(\sqrt{n}+0.155+\frac{0.24}{\sqrt{n}}\right).
\end{equation}
Anomalously large values of $V$ indicate a distribution that is
more clumped than a uniformly random distribution, while low
values mean that angles are more regular. The test statistic is
normalized by the number of variates, $n$, in such a way that
standard tables can be constructed to determine significance
levels for any departure from uniformity; see Fisher (1993). In
this context, however, it is more convenient to determine
significance levels using Monte Carlo simulations of the ``null''
hypothesis of random phases. This is partly because of the large
number of samples available for test, but also because we can use
them to make the test more general.

The first point to mention is that a given set of phases, say
belonging to the modes at fixed $l$ is not strictly speaking
random anyway, because of the constraints noted in the previous
section. One could deal with this by discarding the conjugate
phases, thus reducing the number of data points, but there is no
need to do this when one can instead build the required symmetries
into the Monte Carlo generator.

In addition, suppose the phases of the temperature field over the
celestial sphere were indeed random, but observations were
available only over apart of the sky, such as when a galactic cut
is applied to remove parts of the map contaminated by foregrounds.
In this case the mask may introduce phase correlations into the
observations so the correct null hypothesis would be more
complicated than simple uniform randomness. As long as any such
selection effect were known, it could be built into the Monte
Carlo simulation. One would then need to determine whether $V$
from an observed sky is consistent with having been drawn from the
set of values of $V$ generated over the Monte Carlo ensemble.

There is also a more fundamental problem in applying this test to
spherical harmonic phases. This is that a given set of $a_{l,m}$
depends on the choice of a particular coordinate axis.  A given
sky could actually generate an infinite number of different sets
of $\phi_{l,m}$ because the phase angles are not rotationally
invariant. One has to be sure to take different choices of
$z$-axis into consideration when assessing significance levels, as
a random phase distribution has no preferred axis while systematic
artifacts may. A positive detection of non--randomness may result
from a chance alignment of features with a particular coordinate
axis in the real sky unless this is factored into the Monte Carlo
simulations to. For both the real sky and the Monte Carlo skies we
therefore need not a single value of $V$ but a distribution of
$V$-values obtained by rotating the sky over all possible angles.
A similar approach is taken by Hansen, Marinucci \& Vittorio
(2003). This method may seem somewhat clumsy, but a test is to be
sensitive to departures from statistical homogeneity one should
not base the test on measures that are rotationally invariant,
such as those suggested by Ferreira, Mageuijo \& Gorski (1998) as
these involve averaging over the very fluctuations one is trying
to detect.

\subsection{Rotating  the \protect\( a_{l,m}\protect \)}

In view of the preceding discussion we need to know how to
transform a given set of $a_{l,m}$ into a new set when the
coordinate system is rotated into a different orientation. The
method is fairly standard, but we outline it here to facilitate
implementation of our approach.

Any rotation of the cartesian coordinate system \(
S\{x,y,z\}\mapsto S^{\prime }\{x,y,z\} \) can be described using a
set of three Euler angles $\alpha$, $\beta$, $\gamma$, which
define the magnitude of successive rotations about the coordinate
axes. In terms of a rotation operator $\hat{D}(\alpha, \beta,
\gamma)$, defined so that a field $f(r,\theta,\phi)$ transforms
according to
\begin{equation}
\hat{D}(\alpha, \beta, \gamma) f(r,\theta,\phi) =
f'(r,\theta,\phi)=f(r,\theta', \phi'),
\end{equation}
a vector ${\bf r}$ is transformed as
\begin{equation}
{\bf r}'={\bf D}(0,0,\gamma) {\bf D} (0,\beta,0) {\bf D}(\alpha,
0, 0) {\bf r} \equiv{\bf D}(\alpha, \beta, \gamma) {\bf r}.
\end{equation}
Here ${\bf D}$ is a matrix representing the operator $\hat{D}$,
i.e.
\begin{equation}
{\bf D}(\alpha, \beta, \gamma) = \left( \begin{array}{ccc} \cos
\gamma & \sin \gamma & 0 \\ -\sin \gamma  & \cos \gamma & 0\\ 0 &
0 & 1 \end{array} \right)\left( \begin{array}{ccc} \cos \beta & 0
& -\sin\beta \\ 0  & 1 & 0\\ \sin \beta & 0 & \cos \beta
\end{array} \right)\left( \begin{array}{ccc} \cos
\alpha & \sin \alpha & 0 \\ -\sin \alpha  & \cos \alpha & 0\\ 0 &
0 & 1
\end{array} \right).
\end{equation}
The Wigner \( D \) functions describe the rotation operator used
to realise the transformations of covariant components of tensors
with arbitrary rank  \( l \). The functions, written as \(
D^{l}_{m,m^{\prime }} \), transform a tensor from \( S\{x,y,z\} \)
to \( S^{\prime }\{x^{\prime },y^{\prime },z^{\prime }\} \).
Consider a tensor \( Y_{l,m}(\theta ,\phi ) \) defined under the
coordinate system \( S \) and apply the rotation operator we get:
\begin{equation}
\label{Wigner1} \hat{D}(\alpha ,\beta ,\gamma )Y_{l,m}(\theta
,\phi )=Y_{l,m^{\prime }}(\theta ^{\prime },\phi ^{\prime })=\sum
_{m}Y_{l,m}(\theta ,\phi )D^{l}_{m,m^{\prime }}(\theta ,\phi )
\end{equation}
This means that the transformation of the tensor under the
rotation of the coordinate system can be represented as a matrix
multiplication. Finding the rotated coefficients therefore
requires a simple matrix multiplication once the appropriate \( D
\) function is known. To apply this in practice one needs a fast
and accurate way of generating the matrix elements \(
D^{l}_{m,m^{\prime }} \) for the rotation matrix. There are \(
(2l+1)^{2} \) elements needed to describe the rotation of each
mode and the value of each element depends upon the particular
values of \( (\alpha ,\beta ,\gamma ) \)used for that rotation.
Details of how to implement this are given in Coles et al. (2004).

In order to apply these ideas to make a test of CMB fluctuations,
we first need a temperature map from which we can obtain a
measured set of $a_{l,m}$. Employing the above transformations
with some choice of Euler angles yields a rotated set of the
$a_{l,m}$. It is straightforward to choose a set of angles such
that random orientations of the coordinate axis can be generated.
Once a rotated set has been obtained, Kuiper's statistic is
calculated from the relevant transformed set of phases. For
example, Coles et al. (2004) generated 3000 rotated sets of each
CMB map using this kind of resampling of the original data,
producing 3000 values of $V_{\mathrm{cmb}}$. The values of the
statistic were then binned to form a measured (re-sampled)
distribution of $V_{\mathrm{cmb}}$. The same procedure is applied
to the 1000 Monte Carlo sets of $a_{l,m}$ drawn from a uniformly
random distribution, i.e. each set was rotated 3000 times and a
distribution of $V_{\mathrm{MC}}$ under the null hypothesis is
produced. These realizations were then binned to created an
overall global average distribution under the null hypothesis.

In order to determine whether the distribution of
$V_{\mathrm{cmb}}$ is compatible with a distribution drawn from a
sky with random phases, we use a simple $\chi^2$ test, using
\begin{equation}
\chi^2=\sum_{i}\frac{(f_i-\overline{f_i})^2}{\overline{f_i}}
\end{equation}
where the summation is over all the bins  and $\overline{f_i}$ is
the number expected in the $i$th bin from the overall average
distribution. The larger the value of $\chi^2$ the less likely the
distribution functions are to be drawn from the same parent
distribution. Values of $\chi^2_{\mathrm{MC}}$ are calculated for
the 1000 Monte Carlo distributions and $\chi^2_{\mathrm{cmb}}$ is
calculated from the distribution of $V_{\mathrm{cmb}}$. If the
value of $\chi^2_{\mathrm{cmb}}$ is greater than a fraction $p$ of
the values of $\chi^2_{\mathrm{MC}}$, then the phases depart from
a uniform distribution at significance level $p$. We have chosen
95 per cent as an appropriate level for the level at which the
data are said to display signatures that are not characteristic of
a statistically homogeneous Gaussian random field.

Application of this relatively straightforward method to the WMAP
first-year data shows the existence of phase correlations, as
demonstrated in Figure 2.
\begin{figure}
\centering
\includegraphics[width=0.9\textwidth]{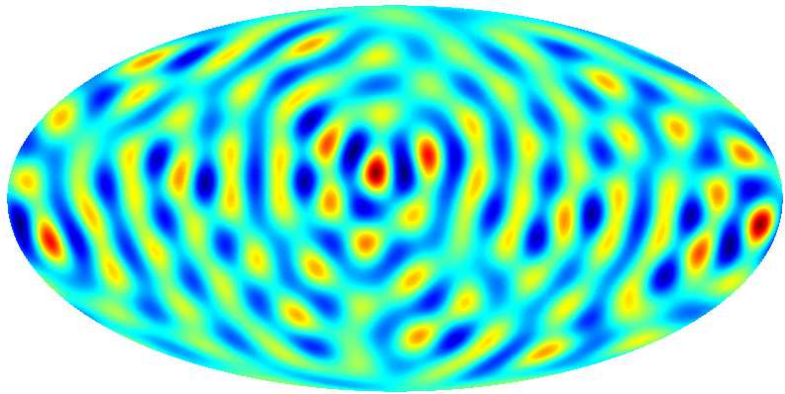}
\caption[]{A reconstruction of the WMAP ILC made using the
spherical harmonic mode amplitudes $a_{l,m}$ for $l=16$ only. Our
analysis method (Coles et al. 2004) shows that these modes at
different $m$ have correlated phases in harmonic space, and the
reconstructed sky shows this is aligned with the Galactic Plane. }
\end{figure}

\subsection{Random Walks in Harmonic Space}

To begin with, we concentrate on a simple measure based on the
distribution of total displacements. Consider a particular value
of $l$. The set of values $\{{\bf a_{l,m}}\}$ can be thought of as
steps in a random walk in the complex plane, a structure which can
be easily visualized and which has well-known statistical
properties.

The simplest statistic one can think of to describe the set
$\{{\bf a_{l,m}}\}$ is the net displacement of a random walk
corresponding to the spherical harmonic mode $l$, i.e.
\begin{equation}
{\bf R_l} = \sum_{m>0} {\bf a_{l,m}},
\end{equation}
where the vector ${\bf a_{l,m}} \equiv (x_{l,m},y_{l,m})$ and the
random walk has an origin at $a_{l,0}$ (which is always on the
$x$-axis). The length of each step $a_{l,m}=|{\bf a_{l,m}}|$ is
the usual spherical harmonic coefficient described in the previous
section and defined by equation (1). If the initial fluctuations
are Gaussian then the two components of each displacement are
independently normal with zero mean and the same variance (8).
Each step then has a Rayleigh distribution so that the probability
density for $a_{l,m}$ to be in the range $(a,a+da)$ is
\begin{equation}
p(a)=\frac{a}{\sigma_l^2}
\exp\left(-\frac{a^2}{2\sigma_l^2}\right).
\end{equation}
This is a particularly simple example of a random walk (McCrea \&
Whipple 1940; Chandrasekhar 1943;  Hughes 1995). Since the
displacements in $x$ and $y$ are independently Gaussian the next
displacement after $l$ steps is itself Gaussian with variance
$l\sigma_l^2$. The probability density of $|R_l|$ to be in the
range $(r,r+dr)$ is then itself a Rayleigh distribution of the
form
\begin{equation}
p_l(r)=\frac{r}{l\sigma_l^2}
\exp\left(-\frac{r^2}{2l\sigma_l^2}\right).
\end{equation}
This requires an estimate of $\sigma_l^2$. This can either be made
using the same data or by assuming a given form for $C_l$, in
which case the resulting test would be of a composite hypothesis
that the fluctuations constitute a Gaussian random field with a
particular spectrum. For large $l$ this is can be done
straightforwardly, but for smaller values the sampling
distribution of $R_l$ will differ significantly from (59) because
of the uncertainty in population variance from a small sample of
$a_{lm}$. This is the so-called ``cosmic variance'' problem.

So far we have concentrated on fixed $l$ with a random walk as a
function of $m$. We could instead have fixed $m$ and considered a
random walk as a function of $l$. Or indeed randomly selected $N$
values of $l$ and $m$. In either case the results above still
stand except with $\sigma_l^2$ replaced by an average over all the
modes considered:
\begin{equation}
\sigma^2=\frac{1}{N} \sum_{l,m} \sigma_{l,m}^2.
\end{equation}
We do not consider this case any further in this paper.

The result (59) only obtains if the steps of the random walk are
independent and Gaussian. If the distribution of the individual
steps is non-Gaussian, but the steps are independent, then the
result (59) will be true for large $l$ by virtue of the Central
Limit Theorem. Exact results for finite $l$ for example
non-Gaussian distributions are given by Hughes (1995). In such
cases the overall 2D random walk comprises two independent 1D
random walks in $x$ and $y$. The Gaussianity of the individual
step components can be tested using their empirical distributions
via a Kolmogorov-Smirnov (K-S) or similar approach. Lack of
independence of step size or step direction (i.e. phase
correlations) would appear as anisotropy of their joint
distribution which could be quantified by direct measures of
cross-correlation or by testing the bivariate distribution using
an appropriate 2D K-S test. The latter task is harder, especially
if the number of modes available is small. Using the net
displacement in 1D or 2D corresponds to using the sum of a sample
of $n$ variables to test the parent distribution. This is not
necessarily powerful, but is robust and has well-defined
properties. The true advantage of the random-walk representation
is that it encapsulates the behaviour of the set $\{{\bf
a_{l,m}}\}$ in a graphical fashion which is ideal for data
exploration.

A slightly different approach is to keep each step length
constant. The simplest way of doing this is to define
\begin{equation}
{\bf \hat{R}_l} = \sum_{m>0} \frac{{\bf a_{l,m}}}{|{\bf
a_{l,m}}|},
\end{equation}
so that each step is of unit length but in a random direction.
This is precisely the problem posed in a famous letter by Pearson
(1905) and answered one week later by Rayleigh (1905). In the
limit of large numbers of steps the result maps into the previous
result (59) with $\sigma_l^2=1$ by virtue of the Central Limit
Theorem. For finite values of $l$ there is also an exact result
which can be derived in integral form using a method based on
characteristic functions (Hughes 1995). The result is that the
probability density for ${\bf \hat{R_l}}$ to be in the range
$r,r+dr$ is
\begin{equation}
q_l(r)= r \int_0^{\infty} u J_0(ur)[J_0(u)]^l du.
\end{equation}
The integral is only convergent for $l>2$ but for $l=1$ or $l=2$
straightforward alternative expressions are available (Hughes
1995). One can use this distribution to test for randomness of the
phase angles without regard to the amplitudes.

A simple test of the hypothesis that the fluctuations are drawn
from a statistically homogeneous and isotropic Gaussian random
field on the sky could  be furnished by comparing the empirical
distribution of harmonic random flights with the form (59). As we
explained above, however, the net displacement of the random walk
is a simple but rather crude indication of the properties of the
$\{{\bf a_{l,m}}\}$, as it does not take into account the ordering
of the individual steps.  The possible non-Gaussian behaviour of
the set $\{{\bf a_{l,m}}\}$ is encoded not so much in the net
displacement but in the {\em shape} of the random walk. To put
this another way, there are many possible paths with the same net
displacement, and these will have different shapes depending on
the correlations between step size and direction. Long runs of
directed steps or regular features in the observed structure could
be manifestations of phase correlation (Coles et al. 2004). The
graphical representation of the set $\{{\bf a_{l,m}}\}$ in the
form illustrated by Figure 3 provides an elegant way of
visualizing the behaviour of the harmonic modes and identifying
any oddities. These could be quantified using a variety of
statistical shape measures: moment of inertia (Rudnick, Beldjenna
\& Gaspari 1987), fractal dimension, first-passage statistics,
shape statistics (e.g. Kuhn \& Uson 1982), or any of the methods
use to quantify the shape of minimal spanning trees (Barrow,
Bhavsar \& Sonoda 1985). Specific examples of correlated random
walks are given in Hughes (1995).

\begin{figure}
\centering
\includegraphics[width=0.9\textwidth]{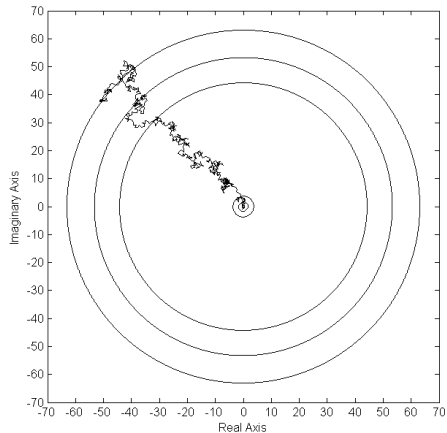}
\caption[]{The random walk performed by the spherical harmonic
coefficients for $l=532$ in the WMAP ILC data, statistically the
mode that displays the greatest departure from that expected under
the null hypothesis. The outer circles correspond to 99.9, 99 and
95 per cent upper confidence limits s (from outer to inner); the
inner circles are the corresponding lower limits, though the 99.9
per cent lower limit is too small to see.}
\end{figure}

In practice the most convenient way to assess the significance of
departures from the relevant distribution would be to perform
Monte Carlo experiments of the null hypothesis. For statistical
measures more complicated than the net displacement, the best way
to set up a statistical test is to use Monte-Carlo re-orderings of
the individual steps to establish the confidence level of any
departure from Gaussianity. This also enables one to incorporate
such complications as galactic cuts.

The WMAP team  released an Internal Linear Combination (ILC) map
that combined  five original frequency band maps in such a way to
maintain unit response to the CMB whilst minimising foreground
contamination. The construction of this map is described in detail
in Bennett et al. (2003). The weighted map is produced by
minimizing the variance of the temperature scale such that the
weights add to one. To further improve the result, the inner
Galactic plane is divided into 11 separate regions and weights
determined separately. This takes account of the spatial
variations in the foreground properties. Thus, the final combined
map does not rely on models of foreground emission and therefore
any systematic or calibration errors of other experiments do not
enter the problem. The final map covers the full-sky and the idea
is that it should represent only the CMB signal. Following the
release of the WMAP 1 yr data Tegmark, Oliveira-Costa \& Hamilton
(2003; TOH) produced a cleaned CMB map. They argued that their
version contained less contamination outside the Galactic plane
compared with the ILC map produced by the WMAP team.

The ILC map is not intended for statistical analysis but in any
case represents a useful ``straw man'' for testing statistical
techniques for robustness. To this end, we analyzed the behaviour
of the random-walks representing spherical harmonic from $l=1$ to
$l=600$ in the WMAP ILC. Similar results are obtained for the TOH
map so we do not discuss the TOH map here. For both
variable-length (57) and unit-length (61) versions of the
random-walk we generated 100000 Monte Carlo skies assuming
Gaussian statistics. These were used to form a distribution of
$|{\bf R_l}|$ (or $|{\bf \hat{R}_l}|$) over the ensemble of
randomly-generated skies. A rejection of the null hypothesis (of
stationary Gaussianity) at the $\alpha$ per cent level occurs when
the measured value of the test statistic lies outstide the range
occupied by $\alpha$ per cent of the random skies.

Application of this simple test to the WMAP data (Stannard \&
Coles 2004) does not strongly falsify the null hypothesis, which
is not surprising given the simplicity of the measure we have
used. The number of modes outside the accepted range is close to
that which would be expected if the null hypothesis were true.
Notice that slightly more modes show up in the unit length case
than in the other, perhaps indicating that the phase correlations
that are known to exist in this data (Chiang et al. 2004) are
masked if amplitude information is also included. The most
discrepant mode turns out to be $l=532$ in both cases. For
interest a plot of the random walk for this case is included as
Figure 3.

\section{Topological Measures of Large-scale Structure}

The application of phase analysis is obviously all performed in
harmonic space (whether Fourier-harmonic or spherical harmonic).
But what does the presence of phase correlations mean for the
morphology of large-scale structure? What is the real-space
morphology of a fluctuation field with random phases? In studying
morphology, one is typically interested in the question of how the
individual filaments, sheets and voids join up and intersect to
form the global pattern shown in Figure 1. Is the pattern
cellular, having isolated voids surrounded by high--density
sheets, or is it more like a sponge in which under-- and
over--dense regions interlock?

Looking at `slice' surveys gives the strong visual impression that
we are dealing with bubbles; pencil beams (deep galaxy redshift
surveys with a narrow field of view, in which the volume sampled
therefore resembles a very narrow cone or ``pencil'') reinforce
this impression by suggesting that a line--of--sight intersects at
more--or--less regular intervals with walls of a cellular pattern.
One must be careful of such impressions, however, because of
elementary topology. Any closed curve in two dimensions must have
an inside and an outside, so that a slice through a sponge--like
distribution will appear to exhibit isolated voids just like a
slice through a cellular pattern. It is important therefore that
we quantify this kind of property using well--defined topological
descriptors.

In an influential series of papers, Gott and collaborators have
developed a method for doing just this (Gott, Melott \& Dickinson
1986; Hamilton, Gott \& Weinberg 1986; Gott et al. 1989; Gott et
al. 1990; Melott 1990; Coles et al. 1996). Briefly, the method
makes use of a topological invariant known as the {\it genus},
related to the {\it Euler--Poincar\'{e} characteristic}, of the
iso--density surfaces of the distribution. To extract this from a
sample, one must first smooth the galaxy distribution with a
filter (usually a Gaussian is used; see \S 14.3) to remove the
discrete nature of the distribution and produce a continuous
density field. By defining a threshold level on the continuous
field, one can construct excursion sets (sets where the field
exceeds the threshold level) for various density levels. An
excursion set will typically consist of a number of regions, some
of which will be simply connected, e.g. a deformed sphere, and
others  which will be multiply connected, e.g. a deformed torus is
doubly connected. If the density threshold is labelled by $\nu$,
the number of standard deviations of the density away from the
mean, then one can construct a graph of the genus of the excursion
sets at $\nu$ as a function of $\nu$: we call this function
$G(\nu)$. The genus can be formally expressed as an integral over
the intrinsic curvature $K$ of the excursion set surfaces,
$S_{\nu}$, by means of the Gauss--Bonnet theorem.

The general form of this theorem applies to any two-dimensional
manifold ${\cal M}$ with any (one--dimensional) boundary $\partial
{\cal M}$ which is piecewise smooth. This latter condition implies
that there are a finite number $n$ vertices in the boundary at
which points it is not differentiable. The Gauss--Bonnet theorem
states that \be \sum_{i=1}^{n} (\pi-\alpha_i) + \int_{\partial
{\cal M}} k_{\rm g} ds + \int_{\cal M} k dA = 2\pi \chi_E({\cal
M}), \ee where the $\alpha_i$ are the angle deficits at the
vertices (the $n$ interior angles at points where the boundary is
not differentiable), $k_{\rm g}$ is the geodesic curvature of the
boundary in between the vertices and $k$ is the Gaussian curvature
of the manifold itself. Clearly $ds$ is an element of length taken
along the boundary and $dA$ is an area element within the manifold
${\cal M}$. The right-hand side of this equation ) is the
Euler--Poincar\'{e} characteristic, $\chi_E$ of the manifold.

This probably seems very abstract but the definition above allows
us to construct useful quantities for both two and
three-dimensional examples. If we have an excursion set as
described above in three-dimensions then its surface can be taken
to define such a manifold. The boundary is just where the
excursion sets intersect the limits of the survey and it will be
taken to be smooth. Ignoring this, we see that the
Euler--Poincar\'{e} characteristic is just the integral of the
Gaussian curvature over the all compact bits of the surface of the
excursion set. Hence, in this case, \be 2\pi\chi_E= \int_{S_{\nu}}
K dS =4\pi\left[ 1- G(\nu) \right]. \ee  Roughly speaking, the
quantity $G$ is the genus, which for a single surface is the
number of ``handles'' the surface posesses; a sphere has no
handles and has zero genus, a torus has one and therefore has a
genus of one. For technical reasons to do with the effect of
boundaries, it has become conventional not to use $G$ but
$G_{S}=G-1$. In terms of this definition, multiply connected
surfaces have $G_S\ge 0$ and simply connected surfaces have
$G_{S}<0$. One usually divides the total genus $G_{S}$ by the
volume of the sample to produce $g_S$, the genus per unit volume.

One of the great advantages of using the genus measure to study
large scale structure, aside from its robustness to errors in the
sample, is that all Gaussian density fields have the same form of
$g_S(\nu)$: \be g_{S}(\nu) = A \Bigl(1-\nu^{2}\Bigr) \exp \Bigl( -
{\nu^{2}\over 2} \Bigr), \ee where $A$ is a spectrum-dependent
normalisation constant. This means that, if one smooths the field
enough to remove the effect of non--linear displacements of galaxy
positions, the genus curve should look Gaussian for any model
evolved from Gaussian initial conditions, regardless of the form
of the initial power spectrum which only enters through the
normalisation factor $A$. This makes it a potentially powerful
test of non--Gaussian initial fluctuations, or of models which
invoke non--gravitational physics to form large--scale structure.
The observations support the interpretation that the initial
conditions were Gaussian, although the distribution looks
non--Gaussian on smaller scales. The nomenclature for the
non--Gaussian distortion one sees is a `meatball shift':
non--linear clustering tends to produce an excess of high--density
simply--connected regions, compared with the Gaussian curve. The
opposite tendency, usually called `swiss--cheese', is to have an
excess of low density simply connected regions in a high density
background, which is what one might expect to see if cosmic
explosions or bubbles formed the large--scale structure. What one
would expect to see in the standard picture of gravitational
instability from Gaussian initial conditions is a `meatball'
topology when the smoothing scale is small, changing to a sponge
as the smoothing scale is increased. This is indeed what seems to
be seen in the observations so there is no evidence of bubbles; an
example is shown in Figure 4.

\begin{figure}
\centering
\includegraphics[width=0.9\textwidth]{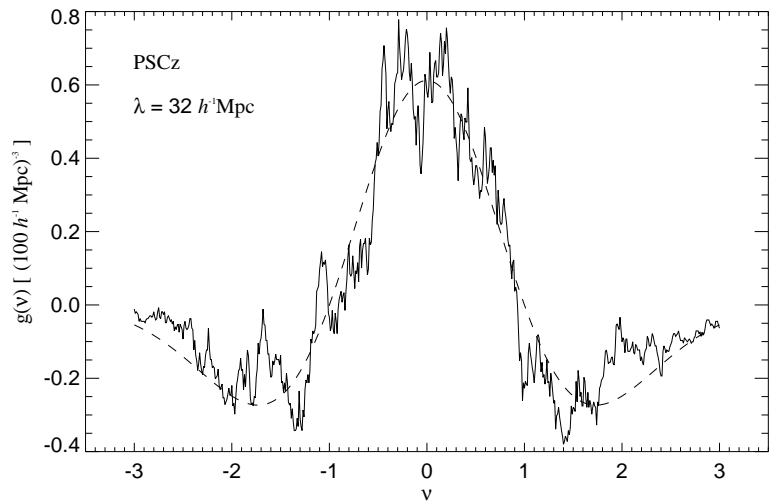}
\caption[]{Genus curve for galaxies in the IRAS PSCz survey. The
noisy curve is the smoothed galaxy distribution while the solid
line is the best--fitting curve for a Gaussian field; from
Canavezes et al. (1998). }\end{figure}

The smoothing required also poses a problem, however, because
present redshift surveys sample space only rather sparsely and one
needs to smooth rather heavily to construct a continuous field. A
smoothing on scales much larger than the scale at which
correlations are significant will tend to produce a Gaussian
distribution by virtue of the central limit theorem. The power of
this method is therefore limited by the smoothing required, which,
in turn, depends on the space--density of galaxies. An example is
given in the Figure, which shows the genus curve for the PSCz
survey of IRAS galaxies.

Topological information can also be obtained from two--dimensional
data sets, whether these are simply projected galaxy positions on
the sky (such as the Lick map, or the APM survey) or `slices'
(such as the various CfA compilations). Here the excursion sets
one deals with are just regions of the plane where the (surface)
density exceeds some threshold. This method can also be applied to
CMB temperature fluctuations where one looks at the topology of
regions bounded by lines of constant temperature (Coles 1988; Gott
et al. 1990; Colley \& Gott 2003; Komatsu et al. 2003).

In such case we imagine the manifold referred to in the statement
of the Gauss--Bonnet theorem to be not the surface of the
excursion set but the surface upon which the set is defined (i.e.
the sky). For reasonably small angles this can be taken to be a
flat plane so that the Gaussian curvature of ${\cal M}$ is
everywhere zero. (The generalization to large angles is trivial;
it just adds a constant curvature term.) The Euler characteristic
is then simply an integral of the line curvature of around the
boundaries of the excursion set: \be 2\pi\chi_E=\int k_{\rm g} ds.
\ee In this case the Euler--Poincar\'{e} characteristic is simply
the number of isolated regions in the excursion set minus the
number of holes in such regions.

This is analogous to the genus, but has the interesting property
that it is an  odd function of $\nu$ for a two--dimensional
Gaussian random field, unlike $G(\nu)$ which is even. In fact the
mean value of $\chi$ per unit area on the sky takes the form \be
\chi(\nu)   = B \nu \exp \Bigl( -\nu^{2}/2 \Bigr), \ee where $B$
is a constant which depends only on the (two--dimensional) power
spectrum of the random field. Notice that $\chi<0$ for $\nu<0$ and
$ \chi>0$ for $\nu>0$. A curve shifted to the left with respect to
this would be a meatball topology, and to the right would be a
swiss--cheese.

There are some subtleties with this. Firstly, as discussed above,
two--dimensional topology does not really distinguish between
`sponge' and `swiss--cheese' alternatives. Indeed, there is no
two-dimensional equivalent of a sponge topology: a slice through a
sponge is topologically equivalent to a slice through
swiss-cheese. Nevertheless, it is possible to assess whether, for
example, the mean density level ($\nu=0$) is dominated by
underdense or overdense regions so that one can distinguish
swiss--cheese and meatball alternatives to some extent.   The most
obviously useful application of this method is to look at
projected catalogues, the main problem being that, if the
catalogue is very deep, each line of sight contains a
superposition of many three--dimensional structures. This
projection acts to suppress departures from Gaussian statistics by
virtue of the central limit theorem. Nevertheless, useful
information is obtainable from projected data simply because of
the  size of the data sets available; as is the case with
three--dimensional studies, the analysis reveals a clear meatball
shift which is what one expects in the gravitational instability
picture. The methods used for the study of two--dimensional galaxy
clustering can also be used to analyze the pattern of fluctuations
on the sky seen in the cosmic microwave background.

More recently, this approach has been generalized to include not
just the Euler--Poincar\'{e} distribution but all possible
topological invariants. This means all quantities that satisfy the
requirement that they be additive, continuous, translation
invariant and rotation invariant. For an excursion set defined in
$d$ dimensions there are $d+1$ such quantities that can be
regarded as independent. Any characteristic satisfying these
invariance properties can be expressed in terms of linear
combinations of these four independent quantities. These are
usually called {\em Minkowski functionals}. Their use in the
analysis of galaxy clustering studies was advocated by Mecke,
Buchert \& Wagner (1994) and has become widespread since then.

In three dimensions there are four Minkowski functionals. One of
these is the integrated Gaussian curvature (equivalent to the
genus we discussed above). Another is the mean curvature, $H$
defined by \be H=\frac{1}{2}\int \left(\frac{1}{R_1} +
\frac{1}{R_2} \right) dA. \ee  In this expression $R_1$ and $R_2$
are the principal radii of curvature at any point in the surface;
the Gaussian curvature is $1/(R_1R_2)$ in terms of these
variables. The other two Minkowski functionals are more
straightforward. They are the surface area of the set and its
volume. These four quantities give a ``complete'' topological
description of the excursion sets.

\section{Discussion}

In this paper I have tried to explain how phase correlations,
arising  from primordial non-Gaussianity, non-linear evolution (or
indeed systematic error) can be measured and use to test
cosmological models. The use of direct phase information is
relatively new in cosmology, so I concentrated on basic properties
and explained in some detail how phases relate to more familiar
descriptors such as the bispectrum and three--point covariance
functions. The magnitude of these statistical descriptors is of
course related to the amplitude of the Fourier modes, but the
factor that determines whether they are zero or non-zero is the
arrangement of the phases of these modes.

The connection between polyspectra and phase information is an
important one and it opens up many lines of future research, such
as how phase correlations relate to redshift distortion and bias.
Using small volumes of course leads to sampling uncertainties
which are  quite straightforward to deal with in the case of the
power-spectra but more problematic for higher-order spectra like
the bispectrum. Understanding the fluctuations about ensemble
averages in terms of phases could also lead to important insights.
On the other hand, the application of phase methods to galaxy
clustering studies is complicated by the non-linear evolution of
perturbations as they collapse and form bound structures.
Structures which are highly localized in real space are highly
dispersed in Fourier space, so it is quite difficult to
disentangle any primordial phase correlations from artifacts of
non-linear evolution.

The CMB is a much more promising arena for the application of
these methods. Late-time non-linear effects should be small (at
least on large angular scales) so any phase correlations will
almost certainly arise from either primordial effects or residual
foreground contamination. The preliminary analysis we have
performed using the WMAP data shows that there are indeed phase
correlations, but Figure 2 suggests the likely interpretation of
this is that it relates to the galaxy. As the constraints on early
Universe physics get stronger, the importance of identifying
low-amplitude foregrounds becomes all the more important. The next
era of CMB physics is likely to be dominated by polarization
studies where the effects of foregrounds are likely to be even
more complicated. There remains a great deal to learn about how to
fully characterize the polarization maps that will soon be
obtained. We can be certain, however, that phase information (in
one way or another) will help us understand what is going on.

%
%

%
%



\printindex
\end{document}